\documentclass[10pt,twoside,twocolumn]{IEEEtran}
\usepackage{graphicx,cite,calc,epstopdf}
\usepackage{tikz}
\usetikzlibrary{shapes,arrows}
\usepackage{rotating}
\usepackage{amsmath,amssymb,amsfonts}
\usepackage{algorithmic}
\usepackage{textcomp}
\ifCLASSINFOpdf
\else
\fi

\begin{document}
%
\title{Dynamic Weight Importance Sampling for Low Cost Spatiotemporal Sensing}


\author{Hadi~Alasti,~\IEEEmembership{Senior Member,~IEEE}
\thanks{H. Alasti is with the School of Polytechnic, Purdue University Fort Wayne (PFW), Fort Wayne, IN, 46805 USA (e-mail: halasti@ieee.org).}
}

\maketitle

\begin{abstract}
A simple and low cost dynamic weight importance sampling (DWIS) implementation is presented and discussed for spatiotemporal sensing of unknown correlated signals in sensor field. The spatial signal is compressed into its contour lines and a partitioned subset of sensors that their observations are in a given margin of the contour levels, is used for importance sampling. The selected sensor population is changed dynamically to maintain the low cost and acceptable spatial signal estimation from limited observations. The estimation performance, cost and convergence of the proposed approach is evaluated for spatial and temporal monitoring, using three different contour level definition schemes. The results show that using DWIS and modeling the spatial signal with contour lines is low cost. In this study the presence of noise in sensor observations is ignored. The number of participant sensors is taken as modeling cost.
\end{abstract}
\begin{IEEEkeywords}
Spatiotemporal monitoring, dynamic weight importance sampling, wireless sensing. 
\end{IEEEkeywords}
\IEEEpeerreviewmaketitle

\section{Introduction}
\IEEEPARstart{I}{n} presentation of spatial signal distributions, modeling using contour lines is a common approach that has been used in variety of applications, such as medical image processing~\cite{Karn}, modeling geometric structures~\cite{Favreau}, etc. This approach is useful in monitoring spatial distributions using wireless sensor networks as it tangibly conserves energy, by compressing the signal distribution into its contour lines~\cite{Hadi_MISS}. Modeling spatial distributions with unequally spaced contours lines has been investigated as an optimal/sub-optimal approach (depends on the reconstruction method) to minimize the modeling error~\cite{Alasti_FOWANC}. Lloyd-Max algorithm~\cite{Sayood}, clearly explains how to calculate the unequally spaced contour levels of a spatial distribution to minimize the modeling error, once the probability density function (pdf) of the signal is known. Unfortunately, pdf is very costly to estimate from the sensor observations, when the spatial signal is unknown. Metropolis-Hastings (MH) algorithm~\cite{Hastings} introduces an approach for estimation of the pdf of a quantity, once sufficiently large number of samples of the quantity are available. In monitoring using wireless sensor network this means spending a good amount of in-network energy, and thus MH is not tractable. Energy conservation is a challenging problem in wireless sensor networks~\cite{Akyildiz}. Sensor selection in sensor network for spatial signal modeling~\cite{Zhang_Selection} is an effective way to conserves the in-network energy. Modeling the spatial signal using contour lines is an example for energy conservative sensor selection.

Importance sampling presents an efficient statistical approach either by sampling another distribution or a subset of samples space from the same distribution for estimation of the statistical properties, such as pdf~\cite{Srinivasan}. The purpose of importance sampling, is to reduce the number of required samples or the sampling cost. However, proposing either a substitute distribution or a subset from the same distribution is the first challenge. Sequential importance sampling (SIS)~\cite{Moral}, which is also known as particle filtering, introduces a practical approach to sequentially filter out a number of samples (sensor observations in sensor network), in order to present an acceptable decision. SIS employs signal processing approaches and Bayesian statistical inference. 

Dynamic weight importance sampling (DWIS)~\cite{F_Liang} is an efficient approach to control the sample population and to maintain the estimation performance. Dynamic weight (DW) employs a variable gain sample-population control mechanism, such as a two state Markov chain with positive or negative variable gains. In each state the sample population is either pruned or enhanced, until convergence to the steady state condition.

In this letter, we employ spatial compression into contour levels, importance sampling and dynamic weight (DWIS) along with signal processing to efficiently monitor an unknown spatial signal using wireless sensing. The uncertainty aspects of the spatial signal are uncertainties at the signal strength range; spatial, spectral and temporal attributes; and its statistical properties.

This letter is organized as follows. In the next section the proposed low cost spatiotemporal sensing algorithm will be introduced. Then, in section III its cost, performance and convergence will be discussed.
\section{System Model and Algorithm}
In this section the system model for spatial signal modeling and then sensor selection based on iterative importance sampling with dynamic weight are detailed. The purpose of this model is to provide a low cost and acceptable estimation of the spatial signal distribution over time. We use the spatial monitoring algorithm that was introduced in~\cite{Hadi_MISS} as the base for the proposed algorithm. Unlike~\cite{Hadi_MISS}, here we do not assume the signal strength range. The proposed algorithm in this letter is performed in two spatial and temporal modeling phases. Also, in this work we assume that during spatial modeling the signal distribution does not change, tangibly.

An unknown spatial signal distribution $g_n(x,y;t)$ is assumed in sensor field with $N$ randomly distributed wireless sensors over the whole field. The spatial signal is modeled with $M$ contour lines at levels $\{\ell_j\}_{j=1}^M$. With this definition, the spatial distribution is compressed into its $M$ contour lines. Because of finite sensor density in the field, those sensors with observations $S_k, \forall k$ that satisfy $\ell_j - \Delta \leq S_k \leq \ell_j + \Delta$ report their observations to the information fusion center (IFC) as contour sensors. 
In each iteration of the spatial signal estimation, the IFC queries the sensor field by sending the $M$ contour levels $\{\ell_j\}_{j=1}^M$ and the contour margin ($\Delta$), and waits for the query replies of the queried sensors. Having received the query replies, the IFC reconstructs an estimation of the spatial distribution. Then, for the next iteration the number of contour levels is incremented i.e., $M \leftarrow M+1$, and the new contour levels are calculated to be introduced to the sensor field.

By increasing the number of contour levels $M$ and keeping $\Delta$ fixed, the total participating sensors (cost) in spatial modeling increases, drastically. To keep it low, three cost reduction mechanisms are used. First, during the spatial modeling each queried sensor reports its observation only once. Second, the number of contour levels $M$ is incremented for a few units i.e., $M \leftarrow M+p$, where $p \geq 2$ and not too big. Third, $\Delta$ changes adaptively in each iteration, related to the spatial estimation error. For the $\Delta$ adaptation mechanism we propose a stochastic gradient approach. 

In calculation of spatial signal estimation error, we use root mean square of error (RMSE), which is calculated at the grid points $(x_i,y_j), i = 1, 2, \cdots P; j = 1, 2, \cdots Q$. To track the convergence of the iterative algorithm, here the tracking RMSE is defined according to (\ref{equ: 1}). In (\ref{equ: 1}), $\tilde{g}_k(x_i,~y_j)$ is the estimation of the spatial signal at grid coordinate point of $(x_i,y_j)$ in the $k$th iteration.
\begin{equation}
Error_k = \sqrt{\sum_{i=1}^P\sum_{j=1}^Q\frac{(\tilde{g}_k(x_i,~y_j)-\tilde{g}_{k-1}(x_i,~y_j))^2}{P\times Q}}
\label{equ: 1}
\end{equation}

As during the iterations the number of contour levels $M$ increases, in a general trend the modeling and tracking error decrease. These decrease is due to finer modeling of the signal with larger $M$, and better spanning of the signal strength range in the process of iterative signal reconstruction and re-sampling. In reconstruction of the spatial signal in each iteration, a larger signal range is discovered from the maximum and the minimum of the new reconstruction that will be used for the contour level range of the next iteration.
Here we define an adaptation algorithm in (\ref{equ: 2}) to dynamically change $\Delta$ related to the slope of the modeling error, in DW process.  
\begin{equation}
	\Delta_k = \Delta_{k-1} (1 + \mu\frac{1}{2\bar{E}_{k-1}}\nabla Error_{k-1})
	\label{equ: 2}
\end{equation}
In~(\ref{equ: 2}), the stochastic gradient of the error function, $\nabla Error_{k-1} = Error_{k-1} - Error_{k-2}$. $\nabla Error_{k-1}$ is normalized by $\bar{E}_{k-1} = (E_{k-1} + E_{k-2})/2$ to make $\Delta$ independent from the instantaneous spatial modeling error value. In~(\ref{equ: 2}), the step-size $\mu$, that its value is in range $0 \leq \mu \leq 1$, trades-off between the error performance and the cost. The smaller $\mu$ results in larger cost and also smaller modeling error, and vice versa.
 
In this study, the contour lines are either equally spaced or unequally spaced, where in the latter case they are calculated based on Lloyd-Max algorithm to minimize the reconstruction error. The Lloyd-Max contour levels are calculated using equations~(\ref{equ: 3}) and ~(\ref{equ: 4})~\cite{Sayood}.
\begin{equation}
\ell_i = \frac{\int_{y_i}^{y_{i+1}}xf_g(x)dx}{\int_{y_i}^{y_{i+1}}f_g(x)dx}, ~~~~i=1,2,\cdots, M
\label{equ: 3}
\end{equation}
where $f_g(x)$ is the pdf of the signal strength, and  $y_i$ is as follows:
\begin{equation}
y_i = \frac{\ell_i + \ell_{i-1}}{2}, ~~~~ i = 1, 2, \cdots, M-1
\label{equ: 4}
\end{equation}
In brief, the spatial distribution is sampled iteratively at location of sensors whose their observations are in $\Delta$ margin of the $M$ contour levels, as subset of all sensors for importance sampling. The $\Delta$ adaptively changes according to~(\ref{equ: 2}) for DW.

For temporal phase, the final $M$ and $\Delta$ in spatial modeling are used, however the signal strength range and the new contour levels are updated for each temporal update. The temporal modeling updates can be done either periodically or based on the local change report of at least $n\%$ of sensors. In this letter the temporal modeling is performed periodically.
\section{Evaluations and Discussion}
The performance of spatial signal estimation over time, its cost and convergence are presented and discussed, in this section. 
We model the spatial signal using diffusion model~\cite{Jindal} due to its simplicity and capability to model correlated spatial distributions. 
The spatial distribution $g(x,y)$ is modeled as the sum of a large number of jointly Gaussian distributions $G(m_{x_i},m_{y_i},\sigma)$ with mean values $m_{x_i}$ and $m_{y_i}$ and standard deviation $\sigma$ and positive known factors $a_i$ and $b_j$, according to (\ref{equ: 5}). In this study, the synthetic spatial signal is created as it is described in (\ref{equ: 5}), with $N_1 = N_2 = 150$, $\sigma_a = 3$ and $\sigma_b = 10$.
\begin{equation}
g(x,y) = \sum_{i=1}^{N_1} a_i G(m_{x_i},m_{y_i},\sigma_a) + \sum_{j=1}^{N_2} b_j G(\acute{m}_{x_j},\acute{m}_{y_j},\sigma_b)
\label{equ: 5}
\end{equation}
Similar to the tracking RMSE, the modeling RMSE is calculated using~(\ref{equ: 6}). 
\begin{equation}
Err_k = \sqrt{\sum_{i=1}^P\sum_{j=1}^Q\frac{(\tilde{g}_k(x_i,~y_j)-g(x_i,~y_j))^2}{P\times Q}}
\label{equ: 6}
\end{equation}

The performance evaluation is done based on computer simulations, using \emph{MATLAB}. The spatial signal is distribution over an area of $100 \times 100$. In this area, 5000 sensors are distributed randomly with uniform distribution. In this letter, we ignore the presence of noise in sensor observations. 
The algorithm is executed in two phases of spatial and temporal modeling. It is assumed that the spatial signal does not tangibly change during each modeling phase. To reduce the spatial modeling cost, it is assumed that each queried sensor replies only once throughout the spatial modeling process.

For modeling the spatial signal, the simulation is performed using three different importance sampling subsets and according to them their related level definitions.
First, the sampling subset is in $\Delta$ neighborhood of equally spaced contour levels with unknown initial signal strength range. The levels related to these subsets is called U-SG as the levels are uniformly spaced and the $\Delta$ margin is adapted using a stochastic gradient form, according to (\ref{equ: 2}). 
Second, the Lloyd-Max levels and their $\Delta$ margin form the sampling subset, with unknown signal strength range and pdf, where $\Delta$ is updated according to (\ref{equ: 2}). We call this levels LM-SG.
The third sampling subset are the sensors that their observations are in $\Delta$ margin of Lloyd-Max levels. Here we assume that the pdf of the signal strength is known and the $\Delta$ is fixed throughout the algorithm, until convergence. We call this scheme LM-fix. 

In each iteration, the IFC passes the sensor observations of the queried sensors to the bi-harmonic spline interpolation re-constructor~\cite{Sandwell} to estimate the $\tilde{g}_k(x,y)$ at grid points of the area. The convergence of the algorithm is studied based on convergence of the tracking RMSE, according to~(\ref{equ: 1}).

For the three introduced level definition schemes, besides the modeling RMSE as estimation performance measure, and the number of query replies from the sensor field as cost, we also investigate the signal strength range estimation for two $\mu$ values.\\
The simulation codes that has been used in this letter is available in~\cite{STM_codes}.
\begin{figure}[t]
		\centering
		\includegraphics[width=0.48\textwidth]{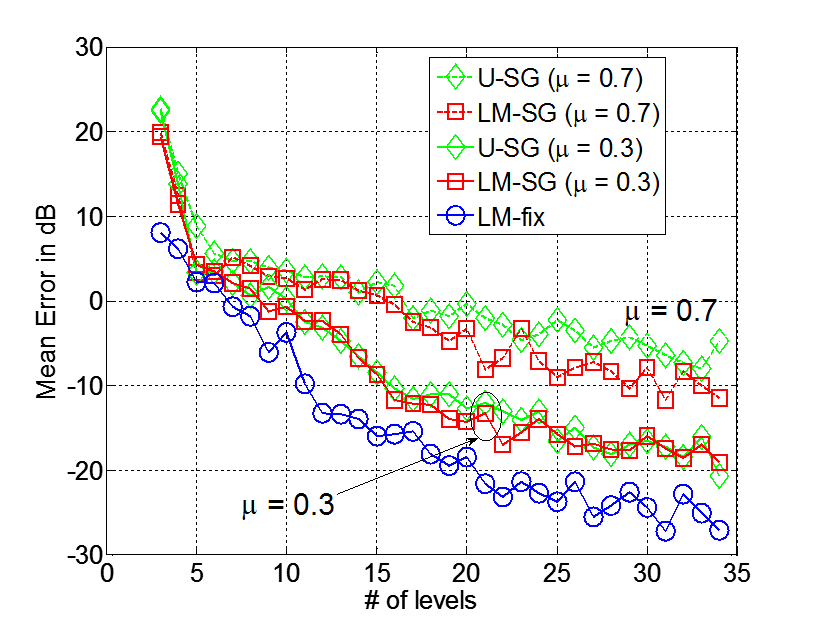}\\
		\caption{RMSE performance of spatial modeling using U-SG, LM-SG and LM-fix for $\mu = 0.3$ and $\mu = 0.7$; $\Delta_0 = 0.2$.}
		\label{fig: 1}
\end{figure}
\begin{figure}[t]
		\centering
		\includegraphics[width=0.48\textwidth]{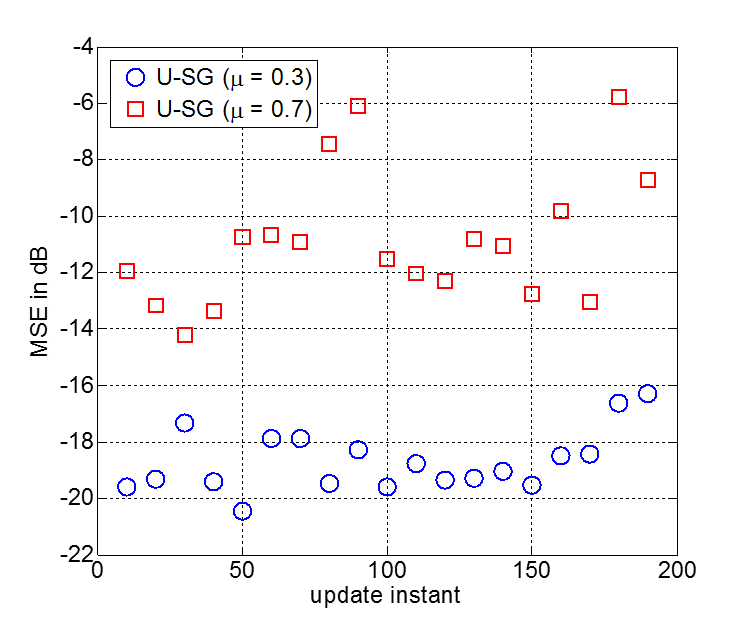}\\
		\caption{RMSE performance of temporal modeling using U-SG, for $\mu = 0.3$ and $\mu = 0.7$.}
		\label{fig: 2}
\end{figure}
\subsection{The Estimation RMSE in Spatial Modeling}
The spatial modeling performance of the DWIS approach for the three introduced contour level schemes are discussed here. The spatial signal is modeled with the observations of sensors that are in $\Delta$ margins of the $M$ contour levels $\{\ell_j\}_{j=1}^M$. This subset of all sensors forms the importance sampling, and the $\Delta$ variations according to (\ref{equ: 2}) adjusts the modeling cost. The higher $\mu$ in (\ref{equ: 2}) results in lower cost and larger modeling error, and vice versa. 
Fig.~\ref{fig: 1}, illustrates the modeling RMSE (according to (\ref{equ: 6})) of DWIS for the three level definition schemes, for $\mu = 0.3$ and $\mu = 0.7$. As the figure shows, by decreasing $\mu$ from $0.7$ to $0.3$, the performance of U-SG and LM-SG becomes closer to that of LM-fix, due to more sensor population. Also, these results show that the performance of U-SG closely tracks LM-SG, where U-SG has lower complexity and needs less processing at IFC.
%

Fig.~\ref{fig: 2}, illustrates the RMSE performance of temporal modeling of DWIS for $0.3$ and $0.7$, only for U-SG, as it has the lowest complexity, and in the absence of perfect pdf it behaves closely similar to LM-SG. The figure shows that the DWIS has relatively steady modeling error in spatial as well as temporal estimation of the signal distribution.
\begin{figure}[t]
		\centering
		\includegraphics[width=0.48\textwidth]{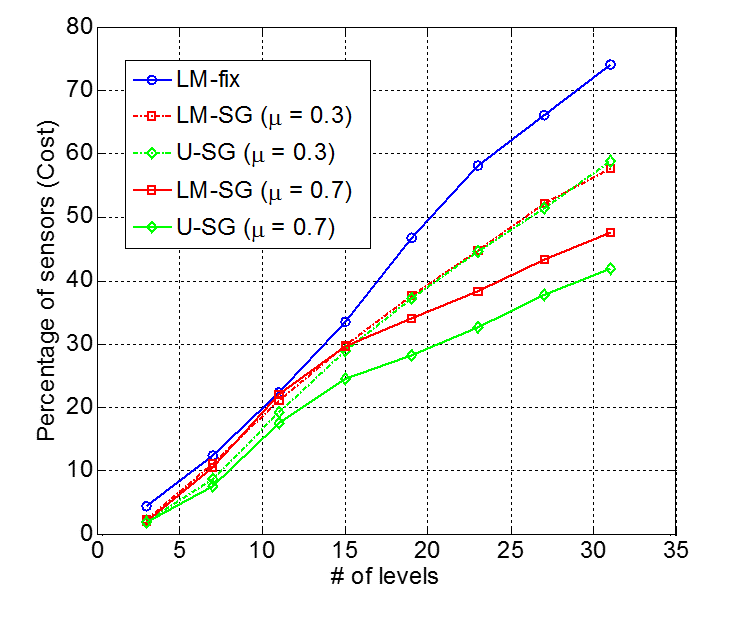}\\
		\caption{The cumulative cost of spatial modeling for U-SG, LM-SG and LM-fix, for $\mu = 0.7$ and $\mu = 0.3$.}
		\label{fig: 3}
\end{figure}
\subsection{The Spatial and Temporal Costs of DWIS}
We define the cumulative cost as the sum of all of the previous iterations' costs of the same process. The cumulative cost is used for spatial modeling as it includes multiple iteration steps.

The spatial modeling's cumulative cost for the 3 contour level definition schemes is shown in Fig.~\ref{fig: 3}, for $\mu = 0.3$ and $\mu = 0.7$. According to these results, U-SG and LM-SG have approximately the same spatial modeling costs for $\mu = 0.3$. The cost of LM-fix is tangibly more than that of U-SG and LM-SG for larger $\mu$ values. Also, as expected by increasing the step-size factor $\mu$, the cost decreases. 
\begin{figure}[t]
		\centering
		\includegraphics[width=0.48\textwidth]{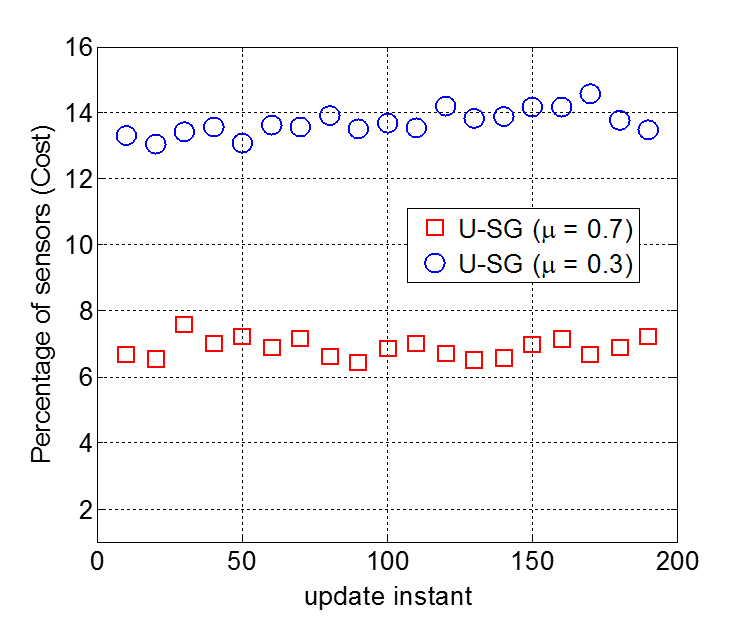}\\
		\caption{The cost of temporal modeling for U-SG with $\mu = 0.3$ and $\mu = 0.7$.}
		\label{fig: 4}
\end{figure}

Fig.~\ref{fig: 4}, compares the cost of temporal modeling for U-SG for two different $\mu$ values. This figure shows that the temporal cost fluctuates for a few deciBells around an average value. Moreover, by increasing the step-size $\mu$, the cost decreases, which is due to reduction in the number of reporting sensor population after convergence to the final value of $\Delta$.
\begin{figure}[h]
		\centering
		\includegraphics[width=0.48\textwidth]{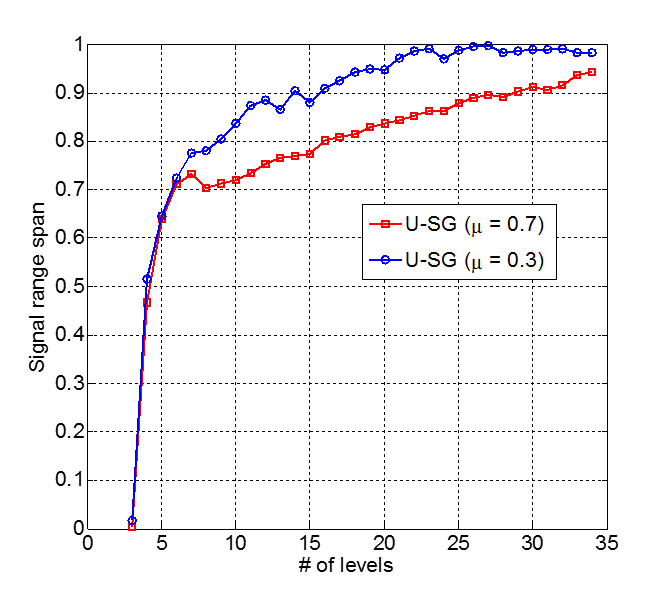}\\
		\caption{Convergence of signal strength range after iterations for $\mu = 0.3$ and $\mu = 0.7$.}
		\label{fig: 5}
\end{figure}
\subsection{Spanning the Signal Strength Range}
One major uncertainty in spatial and temporal modeling of the signal distribution is the unknown signal strength range. The proposed DWIS algorithm, after iterative interpolation and re-sampling of the spatial signal spans its signal strength range. Fig.~\ref{fig: 5}, shows the spanning capability of this algorithm.
\subsection{Convergence of $\Delta$}
\begin{figure}[t]
		\centering
		\includegraphics[width=0.48\textwidth]{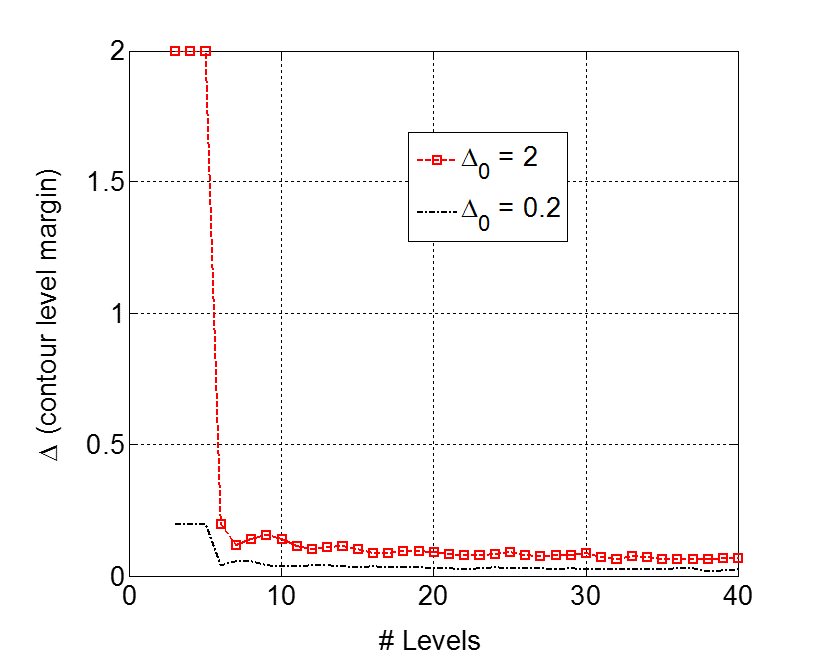}\\
		\caption{Convergence of $\Delta$ for two different initial values $\Delta_0$.}
		\label{fig: 6}
\end{figure}

One important aspect of the proposed DWIS algorithm is its relatively low dependency to the initial $\Delta$ value. Fig.~\ref{fig: 6} shows the convergence of this factor to a tight range for two different initial values of $\Delta_0$. The final $\Delta$ value is a pivotal factor in temporal cost and the modeling RMSE of the algorithm. The final $\Delta$ depends on the spectral attributes of the spatial signal.
\section{Conclusion}
A low cost and tractable dynamic weight importance sampling algorithm is proposed and discussed for spatiotemporal modeling of unknown correlated signals in sensor field. A stochastic gradient formulation with one control factor is proposed to trade-off between the cost and the modeling error. The root mean square error, the number of participant sensors in spatiotemporal modeling, the capability of the algorithm in spanning the signal strength and its convergence are discussed for the proposed algorithm. The results show that the proposed algorithm has low temporal cost and acceptable performance.

\end{document}